\documentclass[sigconf,balance=false]{acmart}
\usepackage{xcolor}
\usepackage{amsmath}
\usepackage{multirow}
\usepackage{graphicx}
\usepackage{subfigure}
\usepackage{colortbl}
\usepackage{makecell}
\usepackage{hyperref}
\hypersetup{
    colorlinks=true,
    linkcolor=blue,
    filecolor=blue,      
    urlcolor=blue,
    citecolor=cyan,
}

\AtBeginDocument{%
  \providecommand\BibTeX{{%
    \normalfont B\kern-0.5em{\scshape i\kern-0.25em b}\kern-0.8em\TeX}}}

\begin{document}
\fancyhead{}

\title{AntM$^{2}$C: A Large Scale Dataset For Multi-Scenario Multi-Modal CTR Prediction}


\settopmatter{authorsperrow=4}
\author{Zhaoxin Huan}
\authornote{These authors contributed equally to this work.}
\author{Ke Ding}
\authornotemark[1]
\affiliation{%
  \institution{Ant Group}
  \city{Hangzhou}
  \country{China}}


\author{Ang Li}
\author{Xiaolu Zhang}
\affiliation{%
  \institution{Ant Group}
  \city{Hangzhou}
  \country{China}}


\author{Xu Min}
\author{Yong He}
\affiliation{%
  \institution{Ant Group}
  \city{Hangzhou}
  \country{China}}


\author{Liang Zhang}
\author{Jun Zhou}
\authornote{Corresponding author.\{zhaoxin.hzx, jun.zhoujun\}@antgroup.com}
\affiliation{%
  \institution{Ant Group}
  \city{Hangzhou}
  \country{China}}

\author{Linjian Mo}
\authornotemark[2]
\author{Jinjie Gu}
\affiliation{%
  \institution{Ant Group}
  \city{Hangzhou}
  \country{China}}


\author{Zhongyi Liu}
\author{Wenliang Zhong}
\affiliation{%
  \institution{Ant Group}
  \city{Hangzhou}
  \country{China}}

\author{Guannan Zhang}
\affiliation{%
  \institution{Ant Group}
  \city{Hangzhou}
  \country{China}}

\renewcommand{\shortauthors}{Zhaoxin Huan et al.}






\begin{abstract}

Click-through rate (CTR) prediction is a crucial issue in recommendation systems, directly impacting user experience and platform revenue.
In recent years, CTR has garnered attention from both industry and academia, leading to the emergence of various public CTR datasets.
However, existing CTR datasets primarily suffer from the following limitations. 
Firstly, users generally click different types of items from multiple scenarios, and modeling the CTR from multiple scenarios can provide a more comprehensive understanding of users and share knowledge between different scenarios. Existing datasets only include CTR data for the same type of items from a single scenario.
Secondly, multi-modal features are essential in multi-scenario CTR prediction as they effectively address the issue of inconsistent ID encoding between different scenarios.
The existing datasets are based on ID features and lack multi-modal features.
Third, a large-scale CTR dataset can provide a more reliable and comprehensive evaluation of complex models, fully reflecting the performance differences between models. While the scale of existing datasets is around 100 million, which is relatively small compared to the real-world industrial CTR prediction.
To address these limitations, we propose AntM$^{2}$C, a \textbf{M}ulti-Scenario \textbf{M}ulti-Modal \textbf{C}TR dataset based on real industrial data from the Alipay platform. Specifically, AntM$^{2}$C possesses the following characteristics: 1) It covers CTR data of 5 different types of items from Alipay, providing insights into the preferences of users for different items, including advertisements, vouchers, mini-programs, contents, and videos. 2) Apart from ID-based features, AntM$^{2}$C also provides 2 multi-modal features, raw text and image features, which can effectively establish connections between items with different IDs.
3) AntM$^{2}$C provides 1 billion CTR data with 200 features, including 200 million users and 6 million items. It is currently the largest-scale CTR dataset available, providing a reliable and comprehensive evaluation for CTR models. Based on AntM$^{2}$C, we construct several typical CTR tasks, including multi-scenario modeling, item and user cold-start modeling, and multi-modal modeling. For each task, we provide comparisons with baseline methods.
The dataset homepage is available at \href{https://www.atecup.cn/home}{https://www.atecup.cn/home}.
\end{abstract}






\maketitle


\section{Introduction}

Click-through rate (CTR) prediction plays a significant role in various domains, including online advertising, search engines, and recommendation systems. CTR prediction refers to the task of estimating the probability that a user will click on a given item. It is essential for optimizing ad revenue, enhancing user experience, and improving engagement. One of the challenging issues in CTR prediction lies in the faithful evaluation of the model.
Public CTR datasets provide a standardized and benchmarked environment for evaluating the performance of different CTR models.
This enables researchers to compare the effectiveness of different models and identify the most suitable ones for specific applications.

However, in order to meet the constantly growing demands of users, the current CTR scenarios and items are becoming increasingly diverse, and the amount of CTR data is also increasing. For example, in Alipay, CTR occurs in the consumer coupons at marketing campaigns, videos on the tab3 page, and mini-programs after a search. As a result, the existing CTR datasets suffer from the following limitations.
Firstly, in real-world industrial CTR prediction, users generally click various types of items from different business scenarios, reflecting their preferences for different items.
For example, on Alipay, a user may browse a video about coffee on the Tab3 page, then click on a coffee coupon during a marketing campaign, and finally use the Alipay search to click a coffee ordering mini-program to place an order.
Jointly modeling this multi-scenario CTR data can provide a more comprehensive understanding of user preferences, and the knowledge across scenarios can be shared to improve the CTR performance in each scenario.
However, existing CTR datasets have a limited range of item types and generally originate from the same business scenario, which fails to capture the multi-scenario preferences of users.
For example, Criteo\footnote{\url{https://www.kaggle.com/c/criteo-display-ad-challenge}} and Avazu\footnote{\url{https://www.kaggle.com/c/avazu-ctr-prediction}} only involve CTR data for advertisements.
As e-commerce platforms, both Amazon\footnote{\url{https://nijianmo.github.io/amazon/index.html}} and AliExpress\footnote{\url{https://tianchi.aliyun.com/dataset/74690}} provide CTR data for their e-commerce items.
Tenrec~\cite{yuan2022tenrec} focuses more on video and article recommendations.
Secondly, multi-modal features can address the issue of inconsistent IDs for similar items in different business scenarios and effectively establish a bridge between different scenarios. For example, a video about coffee and a coffee coupon have different IDs in different business scenarios. Directly using ID features cannot perceive the relationship between these two items. Multi-modal features inherently carry semantic meaning and can better compensate for the inconsistency of ID features across different domains.
Additionally, with the rise of large language models (LLMs), combining LLMs with CTR prediction has become an emerging research field. Existing CTR datasets are based on ID features and lack abundant multi-modal features, resulting in the CTR model being unable to test the performance in multi-scenarios and multi-modal settings.
Furthermore, large-scale datasets can reliably and comprehensively reflect the performance of CTR models, while also highlighting the differences between CTR models.
The existing datasets are typically at the scale of 100 million, which is insufficient to further validate the capabilities in larger-scale industrial scenarios. 

To address the aforementioned challenges, we propose the AntM$^{2}$C dataset, a large-scale multi-scenario multi-modal dataset for CTR prediction. Compared with existing CTR datasets, AntM$^{2}$C has the following advantages:
\begin{itemize}
\item \textbf{Diverse business scenarios and item types}: AntM$^{2}$C contains different types of items from five typical business scenarios on the Alipay platform, including advertisements, vouchers, mini-programs, contents, and videos. Each business scenario has a unique data distribution. The abundant intersecting users and similar items between scenarios enable a more comprehensive evaluation for multi-scenario CTR modeling. Through one evaluation, the effectiveness of the CTR model can be evaluated in multiple business scenarios.

\item \textbf{Multi-modal feature system}: AntM$^{2}$C not only includes ID features but also provides rich multi-modal features such as text and image, which can establish connections between similar items across scenarios and provide better evaluation for multi-modal CTR models. Furthermore, the feature system in AntM$^{2}$C includes up to 200 features\footnote{In the first release, AntM2C open-sourced 10 million samples, including 29 ID features and 2 text features. More data and image features will be gradually released in subsequent phases.}, making it more closely aligned with real-world CTR prediction in industrial scenarios.

\item \textbf{Largest data scale}: AntM$^{2}$C comprises 200 million users and 6 million items, with a total of 1 billion samples\footnotemark[5]. The average number of interactions per user is above 50. To the best of our knowledge, AntM$^{2}$C is the largest public CTR dataset in terms of scale, which can provide comprehensive and reliable CTR evaluation results.

\item \textbf{Comprehensive benchmark}: Based on AntM$^{2}$C, three typical CTR tasks have been built, including multi-scenario modeling, cold-start modeling, and multi-modal modeling. Benchmark evaluation results based on state-of-the-art models are also provided.
\end{itemize}

The rest of the paper is organized as follows. In Section~\ref{sec:related_works}, we briefly review some related works about public CTR datasets. In Section~\ref{sec:data_description}, we give a detailed introduction to the dataset collection and data analysis. In Section~\ref{sec:experiment}, we conduct empirical studies with baseline CTR methods on different CTR tasks.

\begin{figure*}[ht]
\centering\includegraphics[width=1.0\linewidth]{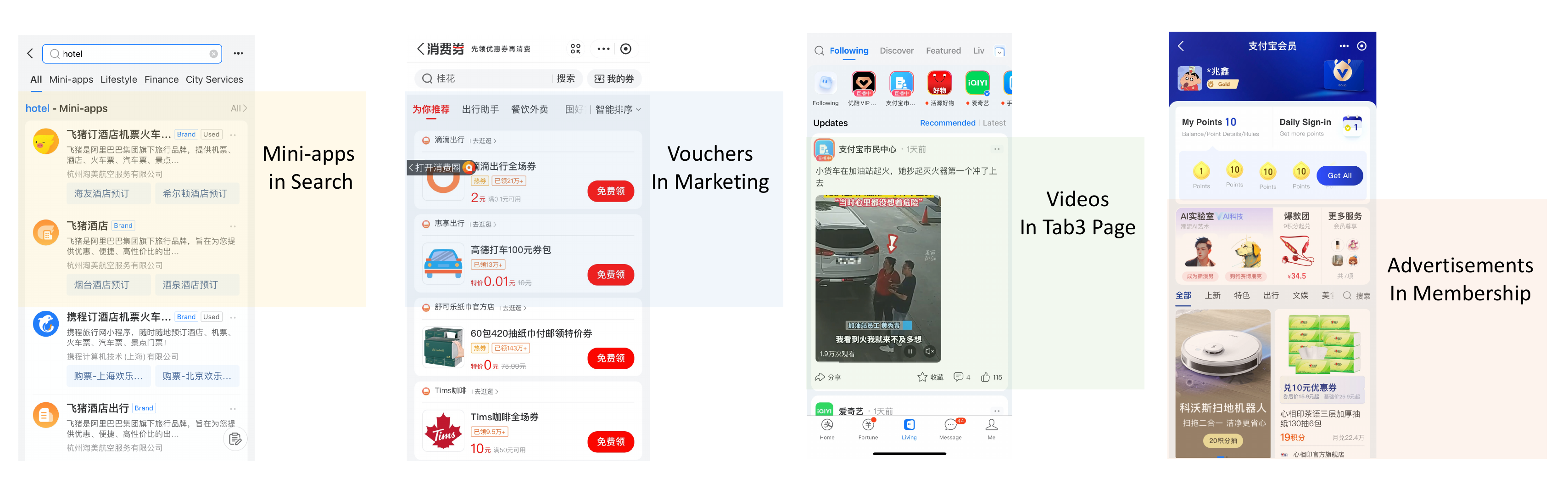}
\caption{An illustration of typical CTR prediction scenarios on the Alipay platform, including service/content search, marketing voucher, Tab3 video recommendation, and advertisement. Each scenario has different types of items, and users have different mindsets when browsing different scenarios.
}
\label{fig:scenes}
\end{figure*}

\section{Existing CTR Datasets}\label{sec:related_works}
The existing public CTR datasets can be roughly divided into two categories: single-scenario and multi-scenario. Both have been widely adopted by the evaluation of CTR methods.

\subsection{Single-Scenario CTR Datasets}
The Criteo dataset is one of the publicly available datasets for CTR prediction. It contains over 45 million records of user interactions with advertisements, including features such as click-through rates, impression rates, and user demographics. Similar to the Criteo dataset, the Avazu dataset contains over 40 million records of user interactions with mobile advertisements. It includes features such as device information, app category, and user demographics. One of the main limitations of the Criteo and Avazu dataset is they only include CTR data for advertisements and cannot be used to evaluate CTR for other business scenarios or types of items. Additionally, the datasets do not provide text information about the advertisement or user, which can limit the scope of the multi-modal modeling.

\subsection{Multi-Scenario CTR Datasets}
The AliExpress is a dataset gathered from real-world traffic logs of the search system in AliExpress. This dataset is collected from 5 countries: Russia, Spain, French, Netherlands, and America, which can be seen as 5 scenarios. It can be used to develop and evaluate CTR prediction models for e-commerce platforms.
The Tenrec dataset is a multipurpose dataset for CTR prediction where click data was collected from two scenarios: articles and videos.
Although the above datasets cover different scenarios, the items within these scenarios are similar. The AliExpress dataset only consists of e-commerce items, and Tenrec involves videos and articles that only reflect the personal interests of users in the entertainment and cultural aspects.
Additionally, similar to single-scenario datasets, both of these datasets lack textual modal information and only provide features such as IDs. This limitation restricts the application of multi-modal modeling.

\section{Data Description}\label{sec:data_description}
\subsection{Data Collection}
AntM$^{2}$C's data is collected from Alipay, a leading platform for payments and digital services. In order to meet the growing demands of users, Alipay recommends various types of items from different business scenarios to users.
\subsubsection{Scenarios}
AntM$^{2}$C collects CTR data in five scenarios on Alipay, and there are differences in the types of items in each scenario.
As shown in Figure~\ref{fig:scenes}, the CTR prediction occurs in multiple scenarios, including services and content on search, vouchers on marketing, videos on Tab3 page, and advertisements on the membership page.
In the search scenario, when a user enters search words, several relevant mini-apps of services or content are displayed for the user to click on.
Marketing scenarios recommend some consumer vouchers, and users click the coupons they are willing to use.
On the Tab3 page, the recommended items are primarily short videos, and users will click to watch the videos they are interested in.
On the membership page, users may click on some online advertisements.
In conclusion, AntM$^{2}$C includes various types of items from different business scenarios. In section~\ref{sec:data_distribution}, we will show that there are differences in the data distribution of these different scenarios. The rich and diverse items provide a more comprehensive evaluation for CTR prediction.

\begin{table}[ht]
\caption{Data statistics of AntM$^{2}$C. To protect user privacy, AntM$^{2}$C anonymizes the scenario names as A-E. The click rate is calculated by dividing the number of clicks by the number of exposures. Since negative sampling is applied to the samples, the click rate may be higher than the actual value.}
\label{tab:statistical}
\resizebox{0.48\textwidth}{!}{
\begin{tabular}{c|ccccc}
\bottomrule
Scenario & Exposure    & Users   & Items   & Click                                   & Click Rate \\ \hline\hline
A        & 3,996,614  & 93,465  & 112,098 &  147,656   & 3.69\%  \\
B        & 8,983,124  & 104,016 & 29,835  & 430,1270  & 47.88\% \\
C        & 1,211,813  & 96,689  & 6,408   & 68,566                                        & 5.66\%  \\
D        & 1,981,484  & 37,095  & 19,092  & 722,009   & 36.44\% \\
E        & 955,162    & 17,904  & 18,265  & 102,671   & 10.75\% \\ \hline
ALL      & 17,128,197 & 120,721 & 184,306 & 5,342,172 & 31.19\% \\ \toprule
\end{tabular}
}
\end{table}

\subsubsection{Data Sampling}
AntM$^{2}$C collects 9-day (from 20230709 to 20230717) CTR samples from the above-mentioned five scenarios and then filters out 1 billion samples of relatively high-activity users who have a total click count $\geq$ 30 across all scenarios.
In the first stage of open sourcing, we randomly sampled 10 million data from these 1 billion samples, and their statistical properties are shown in Table~\ref{tab:statistical}. We will open all 1 billion data in the subsequent stage. For the purpose of protecting user privacy, we do not explicitly indicate the names of the scenarios in the dataset, but instead use the letters 'A-E' as substitutes.

\subsubsection{Data Desensitization}\label{sec:desensitization}
The AntM$^{2}$C does not contain any Personal Identifiable Information (PII) and has been desensitized and encrypted. Each user in the dataset was de-linked from the production system when securely encoded into an anonymized ID.
Adequate data protection measures were undertaken during the experiment to mitigate the risk of data copy leakage.
It is important to note that the dataset is solely utilized for academic research purposes and does not represent any actual commercial use.

\begin{table}[t]
\caption{Overlapped users across the five scenarios in AntM$^{2}$C. AntM$^{2}$C includes the preferences of the same user for items in different scenarios.
}
\label{tab:overlap_users}
\begin{tabular}{c|ccccc}
\bottomrule
Scenario & A                & B                & C                & D                & E                \\ \hline\hline
A        & - & 90537            & 75227            & 19561            & 14937            \\
B        & - & - & 83141            & 22721            & 15978            \\
C        & - & - & - & 31704            & 17019            \\
D        & - & - & - & - & 4788             \\
E        & - & - & - & - & - \\ \toprule
\end{tabular}
\end{table}

\subsection{Data Distribution}

\subsubsection{Data Overlapping}
AntM$^{2}$C contains a portion of overlapped users across the five scenarios. Table~\ref{tab:overlap_users} shows the number of intersecting users among different scenarios, indicating that AntM$^{2}$C can reflect the preferences of the same user for items in different scenarios to effectively conduct multi-scenario CTR evaluation.
As for items, due to the significant diversity in item types among different scenarios, there is no intersection of items between different scenarios.

\begin{figure*}[ht]
\centering\includegraphics[width=0.95\linewidth]{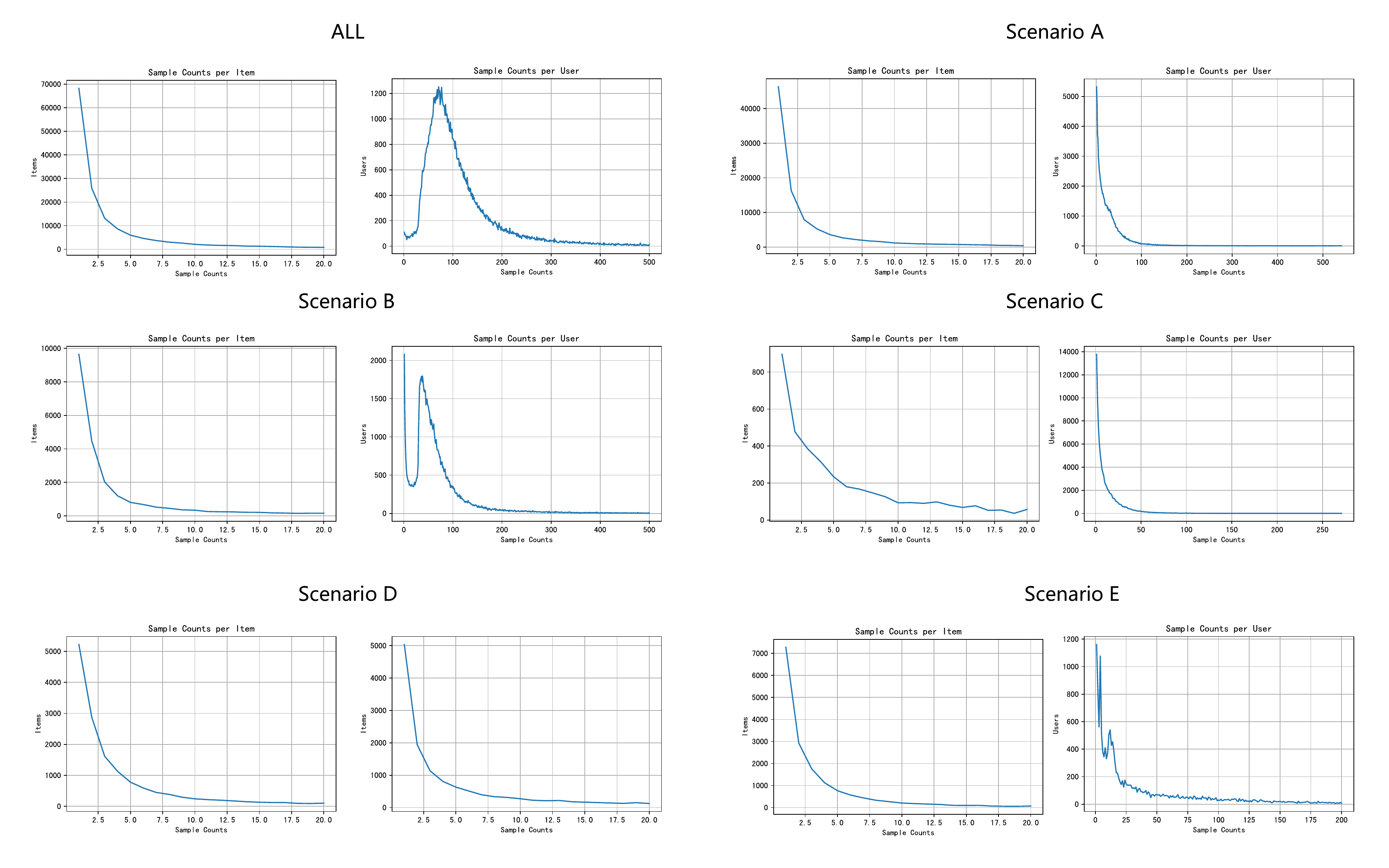}
\caption{An illustration of the data distribution of item (left) and user (right) in different scenarios and the overall samples. The horizontal axis represents exposure frequency, and the vertical axis represents the number of samples at that exposure frequency.
}
\label{fig:distribution}
\end{figure*}

\subsubsection{Item \& User Frequency}\label{sec:data_distribution}
Figure~\ref{fig:distribution} illustrates the frequency of user and item in AntM$^{2}$C dataset, including all samples and samples from different scenarios (A-E). The horizontal axis represents the number of frequencies for users/items, while the vertical axis represents the number of users/items at that frequency. It can be observed that, in terms of item distribution, all scenarios exhibit a long-tail distribution, with 80\% of the sample appearing less than 5 frequencies. This long-tail distribution is consistent with real-world situations.
As for user distribution, there are differences between scenarios. In scenario B, the distribution of user frequency has two peaks, one at less than 5 times and the other around 50 times. After the frequency is greater than 50, the number of users decreases as the frequency increases. In other scenarios, the exposure frequency of users follows a long-tail distribution similar to that of items, where more exposure frequency leads to fewer users. Due to the overlapping users between scenarios, the long-tail distribution of users in multiple scenarios becomes a normal distribution in the global samples. Most users have an exposure frequency of around 50. Overall, the distribution of items and users in AntM$^{2}$C reflects CTR prediction in practice.

\begin{table}[ht]
\caption{Features of AntM$^{2}$C. In addition to ID features, AntM$^{2}$C also includes the raw text features of users and items.}
\label{tab:features}
\resizebox{0.48\textwidth}{!}{
\begin{tabular}{c|l|l|l|c}
\bottomrule
Category                         & Feature\_name              & description                                                     & Type & Coverage             \\ \hline\hline
                                 & user\_id                   & user number                                                     & ID   & 100\%                        \\
                                 & features\_0-26 & user sequences                                         & ID   & 85.50\%                      \\
\multirow{-3}{*}{\begin{tabular}[c]{@{}c@{}}User \\ Features\end{tabular}}  & query\_entity\_seq         & search sequence                                  & Text & 90.32\%                      \\ \hline
                                 & item\_id                   & item number                                                     & ID   & 100\%                        \\
                                 & item\_entity\_names        & entity name of item                      & Text & 100\% \\
\multirow{-3}{*}{\begin{tabular}[c]{@{}c@{}}Item\\ Features\end{tabular}}  & item\_title                & title of item                                                   & Text & 95.50\%                      \\ \hline
                                 & log\_time                  & time in log                & Text & 100\%                        \\
\multirow{-2}{*}{\begin{tabular}[c]{@{}c@{}}Other\\ Features\end{tabular}} & scene                      & scenario number & ID   & 100\%                        \\ \hline
Label                            & label                      & click label                              & Int  & 100\%                        \\ \toprule
\end{tabular}
}
\end{table}

\subsection{Features}
The feature system of AntM$^{2}$C, as shown in Table~\ref{tab:features}, includes ID features of users and items, as well as raw text features.
\subsubsection{User Features}
The user features consist of static profile features\footnote{User static attributes and item title will be open-sourced in the subsequent phases.} and user sequence features.
The static profile features include basic user attributes such as gender, age, occupation, etc.
The sequence features provide the user's recent activities on Alipay, including clicked mini-apps, searched services, purchased items, etc.
\textit{As mentioned in Section~\ref{sec:desensitization}, these user features have been desensitized and encrypted for the purpose of user privacy protection and appear in the dataset in an \textbf{encrypted ID} format, making it impossible to reconstruct the original user features}.
In addition to the ID-based features, AntM$^{2}$C also includes the raw text of user search entities to provide multi-modal evaluation.
\subsubsection{Item Features}
The item features consist of item ID and item textual features. The item ID is a globally unique identifier for each item, and the encoding of item IDs varies across different scenarios. To address the inconsistency of item IDs across scenarios, AntM$^{2}$C also includes the original title text of the items\footnotemark[6] and entities extracted based on the title text.
\subsubsection{Other Features}
In addition to user and item features, AntM$^{2}$C also provides additional features such as log time and scene identification. Users can utilize these extra features to flexibly split the training, validation, and testing sets based on time and evaluate the performance in different scenarios.
\subsubsection{Label}
The label in AntM$^{2}$C indicates whether the user clicked on the corresponding item. If the user clicked, the label is set to 1, otherwise it is set to 0. The ratio of positive to negative samples in AntM$^{2}$C can be obtained from the click rate in Table~\ref{tab:statistical}. It should be noted that there are a large number of negative samples in the actual online logs (samples that were exposed but not clicked on). To address this issue, negative sampling was performed which resulted in a higher click-through rate in the AntM$^{2}$C dataset compared to that in the actual online logs.

\section{Experimental Evaluation}\label{sec:experiment}
In this section, we describe the applications of AntM$^{2}$C in several CTR prediction tasks. We briefly introduce each task and report the results of some baseline methods. 
We select the commonly used AUC (Area Under the Curve) as the metrics for all experiments.
The baseline methods and evaluation results in the experiment provide a demo of using AntM$^{2}$C. More baselines and evaluations will continue to be updated in future work.

\subsection{Multi-Scenario CTR prediction}
Multi-scenario CTR prediction is a common issue in industrial recommendation systems. It builds a unified model by leveraging CTR data from multiple scenarios. The knowledge sharing between scenarios enables the multi-scenario model to achieve better performance compared to single-scene modeling. We conduct an evaluation on multi-scenario CTR prediction using different baseline methods based on the 5 scenarios in the AntM$^{2}$C dataset.

\begin{table}[t]
\caption{The distribution of training and testing data in multi-scenario CTR evaluation. The data is divided by time, and there are differences in the data volume between scenarios.
}
\label{tab:multi_data}
\begin{tabular}{c|c|c}
\bottomrule
Scenario & Train Set  & Test Set  \\ \hline\hline
A        & 3,499,645  & 496,969   \\
B        & 7,890,222  & 1,092,901 \\
C        & 1,059,578  & 151,670   \\
D        & 1,802,707  & 178,777   \\
E        & 846,791    & 104,359   \\ \hline
Total    & 15,098,943 & 2,024,676 \\ \toprule
\end{tabular}
\end{table}

\subsubsection{Data preprocess}\label{sec:multi_data_pre}
In the multi-scenario CTR evaluation, we divide the AntM$^{2}$C dataset based on time, using the data before 20230717 as the training set and the data on 20230717 as the test set. The training and test sets include samples from all five scenarios, and their data distribution is shown in Table~\ref{tab:multi_data}. It can be observed that there are differences in the number of training and test samples among different scenarios. Among them, Scenario B has the highest number of samples, which is ten times that of Scenario E. In terms of features, we use the user and item features from the ID category as shown in Table~\ref{tab:features}. The text features will be used for multi-modal evaluation (see in Section~\ref{sec:multi_modal}).

\subsubsection{Baselines and hyper-parameters}\label{sec:multi_scenarios_baseline}
We mainly choose the multi-task methods as the baseline methods for multi-scenario CTR prediction. We treat the CTR estimation for each scenario as a task and share the knowledge among the scenarios at the bottom layer, with each scenario's CTR score output at the tower layer. The baseline methods and hyperparameter settings are as follows:
\begin{itemize}
    \item DNN: The DNN is trained on a mixture of samples from all scenarios without tasks, serving as the baseline for multi-scenario CTR prediction. The DNN consists of three layers with 128, 32, and 2 units, respectively. The following multi-task model has the same number of layers and unit settings as the DNN.
    \item Shared Bottom~\cite{ruder2017overview}: Shared bottom is the most fundamental model in multi-task learning, where the knowledge is shared among the tasks at the bottom layer. Each task has its own independent tower layer and outputs the corresponding CTR score\footnote{\url{https://github.com/shenweichen/DeepCTR}}.
    \item MMoE~\cite{ma2018modeling}: Based on the shared bottom, MMOE introduces multiple expert networks, each specialized in predicting a specific task, sharing a common input layer. Additionally, MMOE adds a gating network that assigns different weights to each expert based on the input data to determine their influence on predicting the output for a specific task. In the experiment, we set the number of experts in MMOE to 6\footnote{\url{https://github.com/drawbridge/keras-mmoe}}.
    \item PLE~\cite{tang2020progressive}: Based on MMOE, PLE further designs task-specific experts for each task, while retaining the shared expert. This structure allows the model to better learn the differences and correlations among tasks. We set the number of experts in PLE to be the same as MMOE, with each of the five scenarios having its own specific expert and one globally shared expert\footnotemark[7].
\end{itemize}

All baseline methods utilized the Adam~\cite{adam} optimizer with a learning rate of 1e-3 for parameter optimization. The models were trained for 5 epochs with a batch size of 512.

\begin{table}[t]
\caption{Multi-scenario CTR evaluation on AntM$^{2}$C. The table shows the AUC metric of the baseline methods in different scenarios.}
\label{tab:multi_result}
\resizebox{0.48\textwidth}{!}{
\begin{tabular}{c|ccccc}
\bottomrule
\multirow{2}{*}{Methods} & \multicolumn{5}{c}{Scenario}               \\ \cline{2-6} 
                         & A      & B      & C      & D      & E      \\ \hline\hline
DNN                      & 0.7846 & 0.9328 & 0.8733 & 0.6880 & 0.8338 \\
Sharedbottom            & 0.8039 & 0.9414 & 0.8798 & 0.6915 & 0.8525 \\
MMoE                     & 0.7986 & 0.9438 & 0.8751 & 0.6854 & 0.8519 \\
PLE                      & 0.8039 & 0.9429 & 0.8785 & 0.6903 & 0.8506 \\ \toprule
\end{tabular}
}
\end{table}

\subsubsection{Results}
Table~\ref{tab:multi_result} shows the evaluation results of different baseline methods on multi-scenario CTR prediction, from which we can draw the following conclusions. Firstly, compared to the DNN model that trains all data together without considering scenario characteristics, all multi-task models achieve better performance. This demonstrates that in AntM$^{2}$C, there are differences and commonalities between scenarios, and simply mixing training data will not achieve the best results. Secondly, the CTR performance varies across each scenario, indicating different levels of difficulty between scenarios. For example, in scenario B, where there is a large amount of data, the AUC is generally above 0.93, while in scenario D, the AUC is only around 0.68. The diverse business scenarios and items in AntM$^{2}$C enable a more comprehensive and diverse evaluation of CTR. Finally, the expert-structured MMOE and PLE outperform the shared bottom model, demonstrating that refined model design can enhance the performance on AntM$^{2}$C. AntM$^{2}$C is capable of reflecting the differences between different models.

\subsection{Cold-start CTR prediction}
The cold-start problem is a challenging issue in recommendation systems. Training high-quality CTR models using sparse user-item interaction data is a challenging task. Cold-start primarily involves two aspects: users and items. As shown in Figure~\ref{fig:distribution}, the AntM$^{2}$C dataset exhibits a natural long-tail distribution in both users and items. Therefore, we conduct a comprehensive evaluation of cold-start baseline methods based on AntM$^{2}$C dataset.
\subsubsection{Data preprocess}
In cold-start CTR prediction, we split the dataset based on time, using data before 20230717 as the training set and data on 20230717 as the validation and test sets. Based on this data division, we simulated two common cold-start problems in practice: few-shot and zero-shot. 

\begin{itemize}
    \item Few-shot: users and items that appear in the training set with a count greater than 0 and less than $N$\footnote{The selection of this threshold $N$ can vary based on experiments, and we use 100 as an example for all experiments.}, meaning there is only a small amount of training data for these users and items.
    \item Zero-shot: users and items that have never appeared in the training set, indicating that either the user is visiting the scenario for the first time or the item has been launched and added to the scenario on the first day.
\end{itemize}
Table~\ref{tab:cold_data} shows the data distribution of the test set under cold-start CTR evaluation.
By using this dataset division, we can comprehensively evaluate and compare the performance of CTR models on few-shot and zero-shot samples. For few-shot samples, we can observe the model's performance with only a small amount of training data and evaluate the model's generalization ability. For zero-shot samples, we can evaluate the model's recommendation ability on samples that it has never seen before.

\begin{table}[t]
\caption{Data statistics of cold-start CTR evaluation. The meaning of "zero-shot" is that the users and items have never appeared in the training set, while "few-shot" means that there are only a small number of samples of users and items in the training set.
}
\label{tab:cold_data}
\begin{tabular}{c|cc|cc}
\bottomrule
\multirow{2}{*}{Category} & \multicolumn{2}{c|}{Cold-start user} & \multicolumn{2}{c}{Cold-start item} \\ \cline{2-5} 
                          & Count            & Samples           & Count            & Samples          \\ \hline\hline
Few-Shot                  & 67,110           & 685,774           & 30,315           & 306,964          \\
Zero-Shot                 & 65               & 2,752             & 14,230           & 121,447          \\ \toprule
\end{tabular}
\end{table}

\subsubsection{Baselines and hyper-parameters}
The key issue in cold-start modeling is how to learn user preferences and embeddings of users and items with limited data. In recent years, meta-learning-based cold-start methods have become state-of-the-art methods. We selected several representative methods with publicly available code as our baseline models.

\begin{itemize}
    \item DropoutNet~\cite{volkovs2017dropoutnet}: The DropoutNet is a popular cold-start method which applies dropout to control input, and exploits the average representations of interacted items/users to enhance the embeddings of users/items. We implemented the DropoutNet algorithm based on open-source code\footnote{\url{https://github.com/layer6ai-labs/DropoutNet}}.
    
    \item MAML~\cite{finn2017model}: The MAML algorithm is a popular meta-learning approach that aims to enable fast adaptation to new tasks with limited data. MAML learns a good initialization of model parameters that can be effectively adapted to new tasks quickly. We treat each user and item as a task in MAML, and conduct meta-training on warm items. Then we perform meta-testing on cold-start items. The subsequent meta-learning-based algorithms will also follow this task setting.

    \item MeLU~\cite{lee2019melu}: The MeLU algorithm is the first to apply the MAML to address the cold-start problem in recommender systems. Building upon MAML, MeLU ensures the stability of the learning process by not updating the embeddings in the inner loop (support set). The hyperparameter settings in MeLU were determined based on the public code\footnote{\url{https://github.com/hoyeoplee/MeLU}} implementation.

    \item MetaEmb~\cite{pan2019warm}: The MetaEmb algorithm also applies the MAML to address the cold-start problem in recommender systems. Unlike MeLU, MetaEmb focuses on optimizing the embeddings of items. It learns an initial representation using all training samples and then quickly adapts the embeddings of cold-start items. We implemented the MetaEmb algorithm based on open-source code\footnote{\url{https://github.com/Feiyang/MetaEmbedding}}. Although MetaEmb only optimizes the embeddings of items, we have also applied the same approach to optimize the embeddings of users.


\end{itemize}

These base models share the common embedding and DNN structure. The dimensionality of embedding vectors of each input field is fixed to 32 for all our experiments. The Adam optimizer with a learning rate of 1e-3 is used to optimize the model parameters, and the training is performed for 3 epochs with a batch size of 512.
In addition to the aforementioned cold-start algorithms, the DNN (without any cold-start optimization) is also considered as the baseline method for cold-start CTR. 


\begin{table}[t]
\caption{Cold-start evaluation on AntM$^{2}$C. The table shows the AUC metrics of cold start users and items in zero-shot and few-shot situations.
}
\label{tab:cold_result}
\begin{tabular}{c|cc|cc}
\bottomrule
\multirow{2}{*}{Methods} & \multicolumn{2}{c|}{Item} & \multicolumn{2}{c}{User} \\ \cline{2-5} 
                         & Zero-Shot    & Few-Shot   & Zero-Shot   & Few-Shot   \\ \hline\hline
DNN                      & 0.8021       & 0.8339     & 0.7931      & 0.9365     \\
DropNet                  & 0.8097       & 0.8498     & 0.7957      & 0.9387     \\
MAML                     & 0.8131       & 0.8511     & 0.8133      & 0.9393     \\
MeLU                     & 0.8197       & 0.8519     & 0.8103      & 0.9404     \\
MetaEmb                  & 0.8203       & 0.8583     & 0.8091      & 0.9399     \\ \toprule
\end{tabular}
\end{table}

\subsubsection{Results}
Table~\ref{tab:cold_result} shows the CTR performance for cold-start users and items. Because there is limited data for cold start users and items, we do not calculate AUC by scenarios, and evaluate the overall performance of cold start users and items.
From the table, we can observe several phenomena. Firstly, compared to the results shown in Table~\ref{tab:multi_result}, the AUC for cold-start users and items are generally lower than the overall level, which demonstrates that AntM$^{2}$C's data can effectively reflect the differences between cold and warm items and users. Secondly, different cold-start methods show distinguishable results in AntM$^{2}$C, and all of them are significantly better than the DNN model without cold-start optimization. This indicates that AntM$^{2}$C can effectively compare the effects of different cold-start methods and demonstrate the distinctiveness between methods. Finally, the lower performance of zero-shot compared to few-shot indicates that zero-shot CTR prediction is more challenging than few-shot. The two cold start modes provided by AntM$^{2}$C can comprehensively evaluate cold-start CTR prediction.

\subsection{Multi-Modal CTR prediction} \label{sec:multi_modal}
With the rise of large language models (LLMs), it has become a hot research topic to effectively transfer the knowledge of LLM to CTR prediction. There have been many works\cite{sun2019bert4rec,geng2022recommendation,hou2022towards,penha2020does} based on multi-modal CTR modeling using features such as item and user text. AntM$^{2}$C contains raw text features for both users and items, which can provide a more comprehensive evaluation of multi-modal modeling compared to existing CTR datasets. Therefore, we conduct the evaluation of different multi-modal methods based on the AntM$^{2}$C dataset.

\subsubsection{Data preprocess}
In multi-modal evaluation, we adapt the same data processing approach as in multi-scenario evaluation mentioned in Section~\ref{sec:multi_data_pre}, and additionally include the text features from Table~\ref{tab:features}: user query entities and item entities. The text features will be used as inputs to the model together with other ID features.

\subsubsection{Baselines and hyper-parameters}
For the baseline model, we use the language model to process the text features, and then concatenate the text embedding with other ID features and input them into the multi-scenario model described in Section~\ref{sec:multi_scenarios_baseline}. For ease of evaluation, we choose MMoE as the backbone and pre-trained Bert-base\footnote{\url{https://huggingface.co/docs/transformers/main/model_doc/bert}}~\cite{devlin2018bert} as the text embedding extractor. 
The output dimension of Bert's embeddings is 768. Then, a DNN with two layers, each layer having [768, 32] units, is used to reduce the dimension of Bert's embedding to 32. This reduced embedding is concatenated with other features and input into the MMOE model. More powerful language models and the application of text features will continue to be supplemented in future works.

\begin{table}[t]
\caption{Multi-modal evaluation on AntM$^{2}$C. This table shows the AUC metrics for each scenario after incorporating the Bert-base model to model text features based on the multi-task CTR estimation using MMoE.}
\label{tab:multimodal_result}
\begin{tabular}{c|ccccc}
\bottomrule
\multirow{2}{*}{Methods} & \multicolumn{5}{c}{Scenarios}              \\ \cline{2-6} 
                         & A      & B      & C      & D      & E      \\ \hline\hline
MMoE                     & 0.7986 & 0.9438 & 0.8751 & 0.6854 & 0.8519 \\
MMoE+Bert                & 0.7951 & 0.9437 & 0.8851 & 0.6974 & 0.8642 \\ \toprule
\end{tabular}
\end{table}

\subsubsection{Results}
Table~\ref{tab:multimodal_result} shows the evaluation results of the multi-modal CTR. It can be observed that, after adding the text modality, the CTR performance is better in data-sparse scenarios C, D, and E compared to using only the ID modality in the MMoE. Since the current baseline for using the text modality is relatively simple, the improvement in performance is not significant. However, this shows the potential of the text modality provided in AntM$^{2}$C to improve CTR performance.




\section{Conclusion And Future Work}
This paper introduces a large-scale Multi-Scenario Multi-Modal CTR prediction dataset, AntM$^{2}$C dataset.  It includes 1 billion CTR data from five business scenarios on the Alipay platform, and each sample contains multi-modal features in addition to ID features, providing a comprehensive evaluation for CTR models. In the first release, we have made 10 million data publicly available, and we will continue to release more data and features. At the same time, we will gradually evaluate the more state-of-the-art baseline methods on AntM$^{2}$C and provide comprehensive and solid evaluation results.

\bibliographystyle{ACM-Reference-Format}
\bibliography{sigir2024}


\begin{thebibliography}{14}


\ifx \showCODEN    \undefined \def \showCODEN     #1{\unskip}     \fi
\ifx \showDOI      \undefined \def \showDOI       #1{#1}\fi
\ifx \showISBNx    \undefined \def \showISBNx     #1{\unskip}     \fi
\ifx \showISBNxiii \undefined \def \showISBNxiii  #1{\unskip}     \fi
\ifx \showISSN     \undefined \def \showISSN      #1{\unskip}     \fi
\ifx \showLCCN     \undefined \def \showLCCN      #1{\unskip}     \fi
\ifx \shownote     \undefined \def \shownote      #1{#1}          \fi
\ifx \showarticletitle \undefined \def \showarticletitle #1{#1}   \fi
\ifx \showURL      \undefined \def \showURL       {\relax}        \fi
\providecommand\bibfield[2]{#2}
\providecommand\bibinfo[2]{#2}
\providecommand\natexlab[1]{#1}
\providecommand\showeprint[2][]{arXiv:#2}

\bibitem[\protect\citeauthoryear{Devlin, Chang, Lee, and Toutanova}{Devlin
  et~al\mbox{.}}{2018}]%
        {devlin2018bert}
\bibfield{author}{\bibinfo{person}{Jacob Devlin}, \bibinfo{person}{Ming-Wei
  Chang}, \bibinfo{person}{Kenton Lee}, {and} \bibinfo{person}{Kristina
  Toutanova}.} \bibinfo{year}{2018}\natexlab{}.
\newblock \showarticletitle{Bert: Pre-training of deep bidirectional
  transformers for language understanding}.
\newblock \bibinfo{journal}{\emph{arXiv preprint arXiv:1810.04805}}
  (\bibinfo{year}{2018}).
\newblock


\bibitem[\protect\citeauthoryear{Finn, Abbeel, and Levine}{Finn
  et~al\mbox{.}}{2017}]%
        {finn2017model}
\bibfield{author}{\bibinfo{person}{Chelsea Finn}, \bibinfo{person}{Pieter
  Abbeel}, {and} \bibinfo{person}{Sergey Levine}.}
  \bibinfo{year}{2017}\natexlab{}.
\newblock \showarticletitle{Model-agnostic meta-learning for fast adaptation of
  deep networks}. In \bibinfo{booktitle}{\emph{International conference on
  machine learning}}. PMLR, \bibinfo{pages}{1126--1135}.
\newblock


\bibitem[\protect\citeauthoryear{Geng, Liu, Fu, Ge, and Zhang}{Geng
  et~al\mbox{.}}{2022}]%
        {geng2022recommendation}
\bibfield{author}{\bibinfo{person}{Shijie Geng}, \bibinfo{person}{Shuchang
  Liu}, \bibinfo{person}{Zuohui Fu}, \bibinfo{person}{Yingqiang Ge}, {and}
  \bibinfo{person}{Yongfeng Zhang}.} \bibinfo{year}{2022}\natexlab{}.
\newblock \showarticletitle{Recommendation as language processing (rlp): A
  unified pretrain, personalized prompt \& predict paradigm (p5)}. In
  \bibinfo{booktitle}{\emph{Proceedings of the 16th ACM Conference on
  Recommender Systems}}. \bibinfo{pages}{299--315}.
\newblock


\bibitem[\protect\citeauthoryear{Hou, Mu, Zhao, Li, Ding, and Wen}{Hou
  et~al\mbox{.}}{2022}]%
        {hou2022towards}
\bibfield{author}{\bibinfo{person}{Yupeng Hou}, \bibinfo{person}{Shanlei Mu},
  \bibinfo{person}{Wayne~Xin Zhao}, \bibinfo{person}{Yaliang Li},
  \bibinfo{person}{Bolin Ding}, {and} \bibinfo{person}{Ji-Rong Wen}.}
  \bibinfo{year}{2022}\natexlab{}.
\newblock \showarticletitle{Towards universal sequence representation learning
  for recommender systems}. In \bibinfo{booktitle}{\emph{Proceedings of the
  28th ACM SIGKDD Conference on Knowledge Discovery and Data Mining}}.
  \bibinfo{pages}{585--593}.
\newblock


\bibitem[\protect\citeauthoryear{Kingma and Ba}{Kingma and Ba}{2015}]%
        {adam}
\bibfield{author}{\bibinfo{person}{Diederik~P. Kingma} {and}
  \bibinfo{person}{Jimmy Ba}.} \bibinfo{year}{2015}\natexlab{}.
\newblock \showarticletitle{Adam: {A} Method for Stochastic Optimization}. In
  \bibinfo{booktitle}{\emph{3rd International Conference on Learning
  Representations, {ICLR} 2015, San Diego, CA, USA, May 7-9, 2015, Conference
  Track Proceedings}}.
\newblock


\bibitem[\protect\citeauthoryear{Lee, Im, Jang, Cho, and Chung}{Lee
  et~al\mbox{.}}{2019}]%
        {lee2019melu}
\bibfield{author}{\bibinfo{person}{Hoyeop Lee}, \bibinfo{person}{Jinbae Im},
  \bibinfo{person}{Seongwon Jang}, \bibinfo{person}{Hyunsouk Cho}, {and}
  \bibinfo{person}{Sehee Chung}.} \bibinfo{year}{2019}\natexlab{}.
\newblock \showarticletitle{Melu: Meta-learned user preference estimator for
  cold-start recommendation}. In \bibinfo{booktitle}{\emph{Proceedings of the
  25th ACM SIGKDD International Conference on Knowledge Discovery \& Data
  Mining}}. \bibinfo{pages}{1073--1082}.
\newblock


\bibitem[\protect\citeauthoryear{Ma, Zhao, Yi, Chen, Hong, and Chi}{Ma
  et~al\mbox{.}}{2018}]%
        {ma2018modeling}
\bibfield{author}{\bibinfo{person}{Jiaqi Ma}, \bibinfo{person}{Zhe Zhao},
  \bibinfo{person}{Xinyang Yi}, \bibinfo{person}{Jilin Chen},
  \bibinfo{person}{Lichan Hong}, {and} \bibinfo{person}{Ed~H Chi}.}
  \bibinfo{year}{2018}\natexlab{}.
\newblock \showarticletitle{Modeling task relationships in multi-task learning
  with multi-gate mixture-of-experts}. In \bibinfo{booktitle}{\emph{Proceedings
  of the 24th ACM SIGKDD international conference on knowledge discovery \&
  data mining}}. \bibinfo{pages}{1930--1939}.
\newblock


\bibitem[\protect\citeauthoryear{Pan, Li, Ao, Tang, and He}{Pan
  et~al\mbox{.}}{2019}]%
        {pan2019warm}
\bibfield{author}{\bibinfo{person}{Feiyang Pan}, \bibinfo{person}{Shuokai Li},
  \bibinfo{person}{Xiang Ao}, \bibinfo{person}{Pingzhong Tang}, {and}
  \bibinfo{person}{Qing He}.} \bibinfo{year}{2019}\natexlab{}.
\newblock \showarticletitle{Warm up cold-start advertisements: Improving ctr
  predictions via learning to learn id embeddings}. In
  \bibinfo{booktitle}{\emph{Proceedings of the 42nd International ACM SIGIR
  Conference on Research and Development in Information Retrieval}}.
  \bibinfo{pages}{695--704}.
\newblock


\bibitem[\protect\citeauthoryear{Penha and Hauff}{Penha and Hauff}{2020}]%
        {penha2020does}
\bibfield{author}{\bibinfo{person}{Gustavo Penha} {and}
  \bibinfo{person}{Claudia Hauff}.} \bibinfo{year}{2020}\natexlab{}.
\newblock \showarticletitle{What does bert know about books, movies and music?
  probing bert for conversational recommendation}. In
  \bibinfo{booktitle}{\emph{Proceedings of the 14th ACM Conference on
  Recommender Systems}}. \bibinfo{pages}{388--397}.
\newblock


\bibitem[\protect\citeauthoryear{Ruder}{Ruder}{2017}]%
        {ruder2017overview}
\bibfield{author}{\bibinfo{person}{Sebastian Ruder}.}
  \bibinfo{year}{2017}\natexlab{}.
\newblock \showarticletitle{An overview of multi-task learning in deep neural
  networks}.
\newblock \bibinfo{journal}{\emph{arXiv preprint arXiv:1706.05098}}
  (\bibinfo{year}{2017}).
\newblock


\bibitem[\protect\citeauthoryear{Sun, Liu, Wu, Pei, Lin, Ou, and Jiang}{Sun
  et~al\mbox{.}}{2019}]%
        {sun2019bert4rec}
\bibfield{author}{\bibinfo{person}{Fei Sun}, \bibinfo{person}{Jun Liu},
  \bibinfo{person}{Jian Wu}, \bibinfo{person}{Changhua Pei},
  \bibinfo{person}{Xiao Lin}, \bibinfo{person}{Wenwu Ou}, {and}
  \bibinfo{person}{Peng Jiang}.} \bibinfo{year}{2019}\natexlab{}.
\newblock \showarticletitle{BERT4Rec: Sequential recommendation with
  bidirectional encoder representations from transformer}. In
  \bibinfo{booktitle}{\emph{Proceedings of the 28th ACM international
  conference on information and knowledge management}}.
  \bibinfo{pages}{1441--1450}.
\newblock


\bibitem[\protect\citeauthoryear{Tang, Liu, Zhao, and Gong}{Tang
  et~al\mbox{.}}{2020}]%
        {tang2020progressive}
\bibfield{author}{\bibinfo{person}{Hongyan Tang}, \bibinfo{person}{Junning
  Liu}, \bibinfo{person}{Ming Zhao}, {and} \bibinfo{person}{Xudong Gong}.}
  \bibinfo{year}{2020}\natexlab{}.
\newblock \showarticletitle{Progressive layered extraction (ple): A novel
  multi-task learning (mtl) model for personalized recommendations}. In
  \bibinfo{booktitle}{\emph{Proceedings of the 14th ACM Conference on
  Recommender Systems}}. \bibinfo{pages}{269--278}.
\newblock


\bibitem[\protect\citeauthoryear{Volkovs, Yu, and Poutanen}{Volkovs
  et~al\mbox{.}}{2017}]%
        {volkovs2017dropoutnet}
\bibfield{author}{\bibinfo{person}{Maksims Volkovs}, \bibinfo{person}{Guangwei
  Yu}, {and} \bibinfo{person}{Tomi Poutanen}.} \bibinfo{year}{2017}\natexlab{}.
\newblock \showarticletitle{Dropoutnet: Addressing cold start in recommender
  systems}.
\newblock \bibinfo{journal}{\emph{Advances in neural information processing
  systems}}  \bibinfo{volume}{30} (\bibinfo{year}{2017}).
\newblock


\bibitem[\protect\citeauthoryear{Yuan, Yuan, Li, Kong, Li, Chen, Yang, Yu, Hu,
  Li, et~al\mbox{.}}{Yuan et~al\mbox{.}}{2022}]%
        {yuan2022tenrec}
\bibfield{author}{\bibinfo{person}{Guanghu Yuan}, \bibinfo{person}{Fajie Yuan},
  \bibinfo{person}{Yudong Li}, \bibinfo{person}{Beibei Kong},
  \bibinfo{person}{Shujie Li}, \bibinfo{person}{Lei Chen}, \bibinfo{person}{Min
  Yang}, \bibinfo{person}{Chenyun Yu}, \bibinfo{person}{Bo Hu},
  \bibinfo{person}{Zang Li}, {et~al\mbox{.}}} \bibinfo{year}{2022}\natexlab{}.
\newblock \showarticletitle{Tenrec: A Large-scale Multipurpose Benchmark
  Dataset for Recommender Systems}.
\newblock \bibinfo{journal}{\emph{Advances in Neural Information Processing
  Systems}}  \bibinfo{volume}{35} (\bibinfo{year}{2022}),
  \bibinfo{pages}{11480--11493}.
\newblock


\end{thebibliography}

\end{document}